\documentclass[1pt,a4paper]{article}
\usepackage[english]{babel}
\usepackage{color}
\usepackage[english]{babel}
\usepackage{color}
\usepackage{graphicx}
\usepackage{hyperref}
\usepackage{amsmath,amssymb,amsthm}
\usepackage[T1]{fontenc}
\usepackage[utf8]{inputenc}
\usepackage{authblk}
\usepackage[english]{babel}
\usepackage{amsmath,amssymb,amsthm}
\usepackage{color}
\usepackage{graphics}
\usepackage{amsmath,amssymb,amsthm}
\newtheorem{definition}{Definition}

\title{Combining subspace codes with classical linear error-correcting
codes}
\author[1]{Olav Geil}
\author[1]{Louise Foshammer}
\author[1]{Malte Neve-Gr{\ae}sb{\o}ll}
\affil[1]{Department of Mathematical Sciences, Aalborg University}

\begin{document}
\maketitle
\begin{abstract}
{\bf{The below paper was written in May 2014. The authors
    have come to know that there is a significant overlap with
    previous results by Vitaly Skachek, Olgica Milenkovic, and Angelia
    Nedic published in July 2011 as arXiv:1107.4581. Their paper
    entitled ``Hybrid Noncoherent Network Coding'' appeared in {\textit{IEEE Transactions on Information Theory,}} June 2013.}}\\

We discuss how subspace codes can be used to simultaneously correct
errors and erasures when the network performs random linear
network coding and the edges are
noisy channels. This is done by
combining the subspace code with a classical linear error-correcting
code. The classical code then takes care of the errors and the
subspace codes takes care of the erasures.\\

\noindent {\bf{Keywords:}} Linear code, noisy channel, operator channel, random network coding, subspace code. 
\end{abstract}

\section{Introduction}
In this paper we consider networks with one sender and more
receivers. All receivers wish to obtain all messages generated at the
sender. In the seminal paper~\cite{ahlswede} Ahlswede, Cai, Li, and Yeung showed that it is
often possible to obtain a much higher throughput than what can be
achieved by using routing. This is done by allowing the vertices
of the network to linearly combine received information before
forwarding it. This method is now known as linear network
coding. Another important breakthrough was made in~\cite{ho} where it
was shown that if the field size is chosen large enough and if the
coefficients in the linear combinations are chosen by random then with
a very high probability the maximal throughput is attained. The above
model can be extended to also deal with errors and erasures in the
network~\cite{yeung,cai,balli,zhang}. However, another important model
was introduced by K\"{o}tter, Kschischang and Silva in~\cite{kk,sk}
where the network is treated as a black box. They named their channel-model the operator channel and showed how to correct errors and
erasures with respect to the corresponding metric by employing subspace codes. Subspace
codes by now is a very active research area attracting a lot of
attention. In the present paper we are concerned with building a
bridge between the two points of view on a network. We will assume
that what the black box is actually doing is random network coding. It
is well-known \cite{zhang2} that in such a case the strength of subspace codes lies
in its ability to correct erasures and in its ability to keep
adversaries from obtaining too much information~\cite{silvauniversal,kuriharaexplicit} rather than in its
error-correcting ability. In this paper we discuss why subspace codes are
vulnerable to errors in the network. We then show how to combine them
with a classical linear code to obtain simultaneously protection
against errors and erasures.
\section{The protocol for communication through the network}\label{sectwo}\label{secprotocol}
Consider an acyclic network $G=(V,E)$ with one sender
$s$ and $t$ receivers $r_1, \ldots , r_t$. Without loss of generality
we shall assume that $s$ has no incoming edges. All edges in the network have capacity
$1$. We consider a multicast scenario meaning that all receivers wish
to obtain the entire message generated at $s$. As a preparation -- before
sending information into the network -- the given message is encoded
using a subspace code. A subspace code is a collection of subspaces
$C=\{V_1, \ldots , V_{|C|}\}$. Each of these subspaces is called a
code word. We have $V_i \subseteq W \subseteq {\mathbb{F}}_q^k$, $i=1,
\ldots , |C|$, where $W$ is called the ambient space of $C$. When
errors and erasures possibly occurs during the communication process a send
code word will be transformed into another word (vetorspace) in $W$. In the
following we shall without loss of generality always assume that $W$ equals ${\mathbb{F}}_q^k$. We have a set
$M$ of messages with $|M|=|C|$ and a bijective encoding function
identifying each message with a code word (that is, one of the
$V_i$s). We assume that $s$ has $b$ outgoing edges $j_1, \ldots , j_b$ where $b \geq
\max \{ \dim V_1, \ldots , \dim V_{|C|}\}$. When we want to send the message corresponding to $V_i$ we start
by finding by random a generating set $\{\vec{v}_1,\ldots , \vec{v}_b\}
\subseteq {\mathbb{F}}_q^k$ for $V_i$. We then inject the codeword
$V_i$ 
into the network by sending vector $\vec{v}_z$ on edge $j_z$, $z=1,
\ldots , b$. These outgoing edges (channels) may experience errors meaning that
at the receiving end of each edge what arrives is
$Y(j_z)=\vec{v}_{z}+\vec{e}(j_z)$ where $\vec{e}(j_z) \in
{\mathbb{F}}_q^k$ is an error vector. Consider a vertex $u \in V\backslash \{s\}$. Let $i_1,
\ldots , i_l$ be its incoming edges and let $j$ be an outgoing
edge. Denote by $Y(i_1), \ldots , Y(i_l)$ the information arriving at
$u$ along
$i_1, \ldots , i_l$, respectively. Then the information injected into
edge $j$ is $\sum_{s=1}^l f_{i_s,j}Y(i_s)$ where the
coefficients $f_{i_s,j} \in {\mathbb{F}}_q$ are chosen by random. As a
general assumption the coefficients $f_{i,j}$ in the network are
chosen uniformly and independently. Hence, we can assume the encoding
to take place in a distributed manner. At
the receiving end of $j$ what arrives is $Y(j)=\sum_{s=1}^l
f_{i_s,j}Y(i_s)+\vec{e}(j)$, where again $\vec{e}(j)$ is an error
vector. Given a receiver $r$ let $g_1, \ldots , g_w$ be its incoming
edges from which $r$ receives the vectors $Y(g_1), \ldots ,
Y(g_w)$. If the network is noiseless -- meaning that
$\vec{e}(j)=\vec{0}$ for all $j \in E$ -- then we can use the results
from~\cite{ho} to deduce that each receiver $r$ will receive a
generating set for $V_i$ with a probability as close to $1$  as needed provided that the field size $q$ is large enough and that for each
receiver there is a flow from $s$ to $r$ of size at least $\dim
V_i$. For a fixed $q$ there is some fixed probability that things do
not work perfectly and also some of the receivers may not have flows
that are large enough. In this situation what will arrive at receiver
$r$  is a
generating set for some subspace $U$ of $V_i$. The receiver knows
which subspace code that has been used, hence if $U$ is close enough
to $V$ in some meaning that we shall describe in the following section,
the receiver can recover $V_i$. Turning to the
situation where in addition to the above problems also noise is present, receiver $r$ obtains a generating set
for some space $U \subseteq W$. If $U$ is close enough to $V$ in the
meaning described in the next section and if the code $C$ has been chosen
in a clever manner receiver $r$ can recover $V_i$ by a decoding procedure. In
the next section we shall introduce the operator channel which is a
model for communication through networks suitable when subspace codes
are used. Our main concern will be how errors are measured. In the
above model we have the error vectors $\vec{e}(j)$. We shall discuss
how they transform into errors and erasures in the operator channel
model and discuss how to possibly overcome them.
\section{The operator channel}
Following~\cite{kk} we now introduce the operator channel. 
Let $W \subseteq {\mathbb{F}}_q^n$ be an $N$-dimensional
vectorspace (as mentioned in the previous section one will often
assume that $W={\mathbb{F}}_q^k$ and consequently that $N=k$). For an
integer $z \geq 0$ we define a stochastic operator 
${\mathcal{H}}_z$ that given a subspace $V \subseteq W$ returns a
subspace of $V$. If $\dim V > z$ then a randomly chosen
$z$-dimensional subspace is returned. Otherwise, $V$ itself is
returned. 
\begin{definition}\label{defoperator}
An operator channel associated with the ambient space $W$ is a channel
with input and output alphabet ${\mathcal{P}}(W)$ (here
${\mathcal{P}}$ means the set of subspaces of $W$). The channel input $V$ and
channel output $U$ can always be related as $U={\mathcal{H}}_z(V)
\oplus {\mathcal{E}}$ where $z = \dim (U \cap V)$ and ${\mathcal{E}}$ is an error space. We
say that $\rho = \dim V-z$ erasures and $t=\dim {\mathcal{E}}$ errors occurred.
\end{definition} 
We already described the concept of a subspace code. Hence, as the
next thing we introduce a distance measure on ${\mathcal{P}}(W)$ that
matches the definition of erasure and errors in the above definition. 
\begin{definition}
Given $A,B \in {\mathcal{P}}(W)$ the subspace distance $d(A,B)$
is given as $\dim A +\dim B- 2 \dim (A\cap B)$. For a subspace code
the minimum distance is the smallest non-zero distance between codewords.
\end{definition}
By~\cite[Th.\ 2]{kk} a subspace code with minimum distance $d$ allows
for unique decoding whenever the number of erasures and errors sum up
to a number smaller than $d/2$. K\"{o}tter and Kschischang modified
the construction of Gabidulin codes slightly to get a very general
class of codes having very good parameters. Also they devised a
minimum distance decoder for these codes correcting the number of errors and erasures
as described above.\\

In the next section we shall discuss how errors and erasures can occur
when the protocol of Section~\ref{sectwo} is applied. 
\section{Errors and erasures}
For a receiver $r$, a cut in the network is a partition of $V$ into two
disjoint sets $P_1$ and $P_2$ such that $s \in P_1$ and $r \in
P_2$. The corresponding edges from $P_1$ to $P_2$ are denoted by
$C(P_1,P_2)$. Cuts play a crucial role in network coding. We start
this section by discussing a claim which at a first glance seems
correct, but which is actually wrong, namely that an error-vector
$\vec{e}(j)$ which is different from $\vec{0}$ may cause widespread
error propagation. Consider a receiver $r$ and a corresponding cut
such that $C(P_1,P_2)$ contains $j$ (if such a cut does not exist then the
error does not have any implication for what is received on the
ingoing edges to $r$). If we name the edges in $C(P_1,P_2)$ different
from $j$ by
$\ell_1, \ldots , \ell_x$ then $X={\mbox{Span}} \{Y(\ell_1), \ldots
, Y(\ell_x),Y(j)\}$ is what is passed on from the part of the network
containing $P_1$ to the remaining part of the network. If in a
``later'' cut $P_1^\prime,P_2^\prime$ with $P_1 \subsetneq P_1^\prime$
the $Y$-values traveling on $C(P_1^\prime,P_2^\prime)$ does not span
$X$ but spans some other space $X^\prime$ then it is solely due to
addition of new errors or/and it is a consequence of dimension loss caused
by a bad choice of coding coefficients $f_{s,t}$ or by the lack of a
flow of size at least the dimension of $\dim X$. Hence, errors do not
propagate.\\

The correct reason that subcodes are often not very good when the
network experience noise is that altering even a single of the symbols send
into an edge will often cause one error as well as 
one  
erasure, Definition~\ref{defoperator}. (In principle an
error vector $\vec{e}(j)$ may cause $-1$, $0$ or $1$ error and $-1$,
$0$ or $1$ erasure.) In many cases such errors and erasures caused by different
edges will not cancel out each other.
Hence if a non-trivial portion of the edges $E$ experience noise it
may in total have a dramatic effect. A reasonable model would be to
assume that the edges are all $q$-ary symmetric channels with the same
error probability. If this probability is not extremely small and if
the vectors of $W$ are not very short then a network with even a
modest number of edges will cause many errors and erasures to the code
word $V_i$. In the following we shall always assume that the edges
correspond to memoryless channels with an identical probability $p$ of
errors and probability $0$ for erasure. These channels are always
assumed to act independently
of each other. \\

We conclude the section by mentioning that the lack of flows of size equal to $\dim V_i$
and bad choices of coding coefficients $f_{i,j}$ can of course cause
erasures. But this should happen with a low probability if the maximum
dimension of code words in $C$ are not too high compared to the
expected min cut of the network.
\section{Using an additional linear code}
Fortunately, there is a simple fix to the problem of errors that we 
described in the previous section. Consider as in Section~\ref{secprotocol}
a generating set of vectors $\{ \vec{v}_1, \ldots, \vec{v}_b\}$ for
$V_i$. These vectors are of length $k$. Now as a preparation before
sending them on the outgoing edges of $s$ we protect them by a, say
systematic, linear
code $D$ with parameters $[n,k,\delta]$ and obtain a new set of
vectors $\{\vec{c}_1, \ldots , \vec{c}_b\}\subseteq D \subseteq
{\mathbb{F}}_q^n$. These are the vectors send on the outgoing edges of
$s$. At any point of the communication
the $k$ first symbols are unaffected by this action. However, any
linear combination of $\vec{c}_1, \ldots , \vec{c}_b$ is still in
$D$. Hence, a receiver $r$ can simply start by performing for each of
its incoming edges the decoding
algorithm of $D$ to the incoming vector in ${\mathbb{F}}_q^n$. If the
minimum distance $\delta$ of $D$ are large enough the first $k$
symbols of the resulting
vectors will with high probability span a subspace of $V_i$, meaning
that we have corrected the errors. To hopefully recover $V_i$ from this
subspace we finally perform the decoding algorithm of the subspace
code $C$.\\

Recall, that we previously assumed that all edges correspond to
memoryless channels with error probability $p$ and erasure probability
$0$. Also recall that these channels are assumed to act independently
of each other. For an incoming edge $i$ to a receiver $r$ we define
\begin{eqnarray}
K(i)&=&\# \{ e \in E \mid e {\mbox{ belongs to a path from }} s {\mbox{ to
  }} r \nonumber \\
&&{\mbox{ with the last edge being }} i\}. \nonumber
\end{eqnarray}
Define 
$$K=\max \{ K(i) \mid i {\mbox{ is an incoming edge for some
    receiver }} r\}.$$
We may have some information on the topology of the network allowing
us to derive an upper estimate $K \leq K^\prime$. The linear code $D$
then should be chosen such that it is suitable for a channel with
probability for errors being 
$$1- (1-p)^{K^\prime}$$
and the probability of erasure being $0$.\\

The above method of course comes with the price of a drop in
communication rate. We leave it as an open research problem if
possibly the two error-corrections involved could be integrated with
each other in such a way that the drop in communication rate is
less dramatic.
\section{Concluding remarks}
It is known that if one protects the communication on each
edge by a linear code and that if one performs a decoding algorithm at
the end point of that edge then one can attain the capacity of the
network if simultaneously one use linear network coding \cite{song2006separation,zhang}. Our
approach is somehow related as it also treats error-correction and
erasure correction independently of each other. Unfortunately, our
approach does not attain the capacity of the network. However, there are
situations where subspace codes are natural to use, and in such
situations our approach provides a procedure to deal with noise.
\section*{Acknowledgments}
The authors gratefully acknowledge the support from
The Danish Council for Independent Research (Grant
No.\ DFF--4002-00367). They also would like to thank Muriel M\'{e}dard, Frank Fitzek, Diego
Ruano, 
Daniel Lucani Roetter, and Morten Videb{\ae}k Pedersen for
enlightening discussions.


\begin{thebibliography}{10}

\bibitem{ahlswede}
Ahlswede, R., Cai, N., Li, S.Y., Yeung, R.W.: Network information flow.
\newblock Information Theory, IEEE Transactions on \textbf{46}(4), 1204--1216
  (2000)

\bibitem{balli}
Balli, H., Yan, X., Zhang, Z.: On randomized linear network codes and their
  error correction capabilities.
\newblock Information Theory, IEEE Transactions on \textbf{55}(7), 3148--3160
  (2009)

\bibitem{cai}
Cai, N., Yeung, R.W., et~al.: Network error correction, ii: Lower bounds.
\newblock Communications in Information \& Systems \textbf{6}(1), 37--54 (2006)

\bibitem{ho}
Ho, T., M{\'e}dard, M., Koetter, R., Karger, D.R., Effros, M., Shi, J., Leong,
  B.: A random linear network coding approach to multicast.
\newblock Information Theory, IEEE Transactions on \textbf{52}(10), 4413--4430
  (2006)

\bibitem{kk}
Koetter, R., Kschischang, F.R.: Coding for errors and erasures in random
  network coding.
\newblock Information Theory, IEEE Transactions on \textbf{54}(8), 3579--3591
  (2008)

\bibitem{kuriharaexplicit}
Kurihara, J., Uyematsu, T., Matsumoto, R.: Explicit construction of universal
  strongly secure network coding via mrd codes.
\newblock In: Information Theory Proceedings (ISIT), 2012 IEEE International
  Symposium on, pp. 1483--1487. IEEE (2012)

\bibitem{silvauniversal}
Silva, D., Kschischang, F.R.: Universal secure network coding via rank-metric
  codes.
\newblock Information Theory, IEEE Transactions on \textbf{57}(2), 1124--1135
  (2011)

\bibitem{sk}
Silva, D., Kschischang, F.R., Koetter, R.: A rank-metric approach to error
  control in random network coding.
\newblock Information Theory, IEEE Transactions on \textbf{54}(9), 3951--3967
  (2008)

\bibitem{song2006separation}
Song, L., Yeung, R.W., Cai, N.: A separation theorem for single-source network
  coding.
\newblock Information Theory, IEEE Transactions on \textbf{52}(5), 1861--1871
  (2006)

\bibitem{yeung}
Yeung, R.W., Cai, N., et~al.: Network error correction, i: Basic concepts and
  upper bounds.
\newblock Communications in Information \& Systems \textbf{6}(1), 19--35 (2006)

\bibitem{zhang}
Zhang, Z.: Linear network error correction codes in packet networks.
\newblock Information Theory, IEEE Transactions on \textbf{54}(1), 209--218
  (2008)

\bibitem{zhang2}
Zhang, Z.: Theory and applications of network error correction coding.
\newblock Proceedings of the IEEE \textbf{99}(3), 406--420 (2011)

\end{thebibliography}
\end{document}